\documentclass[10pt,twocolumn,twoside]{IEEEtran}

\usepackage{amssymb}
\usepackage{amsfonts}
\usepackage{amsmath}
\usepackage{cite}
\usepackage[dvips]{graphicx}
\usepackage{color}
\usepackage{comment}
\usepackage{makecell}

\begin{document}
\newtheorem{definition}{\it Definition}
\newtheorem{theorem}{\bf Theorem}
\newtheorem{lemma}{\it Lemma}
\newtheorem{corollary}{\it Corollary}
\newtheorem{remark}{\it Remark}
\newtheorem{example}{\it Example}
\newtheorem{case}{\bf Case Study}
\newtheorem{assumption}{\it Assumption}
\newtheorem{property}{\it Property}
\newtheorem{proposition}{\it Proposition}

\newcommand{\hP}[1]{{\boldsymbol h}_{{#1}{\bullet}}}
\newcommand{\hS}[1]{{\boldsymbol h}_{{\bullet}{#1}}}

\newcommand{\ba}{\boldsymbol{a}}
\newcommand{\baq}{\overline{q}}
\newcommand{\bA}{\boldsymbol{A}}
\newcommand{\bb}{\boldsymbol{b}}
\newcommand{\bB}{\boldsymbol{B}}
\newcommand{\bc}{\boldsymbol{c}}
\newcommand{\bcO}{\boldsymbol{\cal O}}
\newcommand{\bh}{\boldsymbol{h}}
\newcommand{\bH}{\boldsymbol{H}}
\newcommand{\bl}{\boldsymbol{l}}
\newcommand{\bm}{\boldsymbol{m}}
\newcommand{\bn}{\boldsymbol{n}}
\newcommand{\bo}{\boldsymbol{o}}
\newcommand{\bO}{\boldsymbol{O}}
\newcommand{\bp}{\boldsymbol{p}}
\newcommand{\bq}{\boldsymbol{q}}
\newcommand{\bR}{\boldsymbol{R}}
\newcommand{\bs}{\boldsymbol{s}}
\newcommand{\bS}{\boldsymbol{S}}
\newcommand{\bT}{\boldsymbol{T}}
\newcommand{\bw}{\boldsymbol{w}}

\newcommand{\balpha}{\boldsymbol{\alpha}}
\newcommand{\bbeta}{\boldsymbol{\beta}}
\newcommand{\bOmega}{\boldsymbol{\Omega}}
\newcommand{\bTheta}{\boldsymbol{\Theta}}
\newcommand{\bphi}{\boldsymbol{\phi}}
\newcommand{\btheta}{\boldsymbol{\theta}}
\newcommand{\bvarpi}{\boldsymbol{\varpi}}
\newcommand{\bpi}{\boldsymbol{\pi}}
\newcommand{\bpsi}{\boldsymbol{\psi}}
\newcommand{\bxi}{\boldsymbol{\xi}}
\newcommand{\bx}{\boldsymbol{x}}
\newcommand{\by}{\boldsymbol{y}}

\newcommand{\cA}{{\cal A}}
\newcommand{\bcA}{\boldsymbol{\cal A}}
\newcommand{\cB}{{\cal B}}
\newcommand{\cE}{{\cal E}}
\newcommand{\cG}{{\cal G}}
\newcommand{\cH}{{\cal H}}
\newcommand{\bcH}{\boldsymbol {\cal H}}
\newcommand{\cK}{{\cal K}}
\newcommand{\cO}{{\cal O}}
\newcommand{\cR}{{\cal R}}
\newcommand{\cS}{{\cal S}}
\newcommand{\dcS}{\ddot{{\cal S}}}
\newcommand{\ds}{\ddot{{s}}}
\newcommand{\cT}{{\cal T}}
\newcommand{\cU}{{\cal U}}
\newcommand{\wt}[1]{\widetilde{#1}}

\newcommand{\mA}{\mathbb{A}}
\newcommand{\mE}{\mathbb{E}}
\newcommand{\mG}{\mathbb{G}}
\newcommand{\mR}{\mathbb{R}}
\newcommand{\mS}{\mathbb{S}}
\newcommand{\mU}{\mathbb{U}}
\newcommand{\mV}{\mathbb{V}}
\newcommand{\mW}{\mathbb{W}}

\newcommand{\uq}{\underline{q}}
\newcommand{\ubq}{\underline{\boldsymbol q}}

\newcommand{\red}[1]{\textcolor[rgb]{1,0,0}{#1}}
\newcommand{\gre}[1]{\textcolor[rgb]{0,1,0}{#1}}
\newcommand{\blu}[1]{\textcolor[rgb]{0,0,1}{#1}}

\title{Towards Ubiquitous AI in 6G with Federated Learning} 

\author{Yong~Xiao, 
Guangming~Shi, 
and Marwan~Krunz 

\thanks{Y. Xiao is with the School of Electronic Information and Communications at the Huazhong University of Science and Technology, Wuhan, China 430074 (e-mail: yongxiao@hust.edu.cn).

G. Shi is with the School of Artificial Intelligence, the Xidian University, Xi'an, Shaanxi 710071, China (e-mail: gmshi@xidian.edu.cn),

M. Krunz is with the Department of Electrical and Computer Engineering at the University of Arizona, Tucson, AZ 85710 (e-mail: krunz@email.arizona.edu).
}
}

\maketitle

\begin{abstract}
With 5G cellular systems being actively deployed worldwide, the research community has started to explore novel technological advances for the subsequent generation, i.e., 6G. It is commonly believed that 6G will be built on a new vision of ubiquitous AI, an hyper-flexible architecture that brings human-like intelligence into every aspect of networking systems. Despite its great promise, there are several novel challenges expected to arise in ubiquitous AI-based 6G. Although numerous attempts have been made to apply AI to wireless networks, these attempts have not yet seen any large-scale implementation in practical systems. One of the key challenges is the difficulty to implement distributed AI across a massive number of heterogeneous devices. Federated learning (FL) is an emerging distributed AI solution that enables data-driven AI solutions in heterogeneous and potentially massive-scale networks. Although it still in an early stage of development, FL-inspired architecture has been recognized as one of the most promising solutions to fulfill ubiquitous AI in 6G. In this article, we identify the requirements that will drive convergence between 6G and AI. We propose an FL-based network architecture and discuss its potential for addressing some of the novel challenges expected in 6G. Future trends and key research problems for FL-enabled 6G are also discussed.
\end{abstract}

\section{Introduction}
\label{Section_Introduction}
The field of wireless communications has evolved rapidly over the past few decades, driven largely by growing demand for mobile Internet and wireless-enabled applications. Third Generation (3G) technology enabled wireless data services for texting and basic web browsing, Fourth Generation (4G) systems popularized mobile video streaming, and the most recent Fifth Generation (5G) systems are designed to support augmented reality/virtual reality (AR/VR), massive-scale Internet-of-Things (IoT), and autonomous vehicles, among others. With 5G being actively deployed on a global scale, the research community has started to look into the next-generation wireless technology, i.e., Sixth Generation (6G).

5G systems represented a major paradigm shift from a communication-oriented architecture to a service-based architecture (SBA), enabling a plethora of applications across diverse verticals. 6G systems are expected to take a substantially more holistic approach, catalyzing innovative applications and smart services by performing timely and highly efficient data collection, transportation, learning, and synthesizing anywhere at any time. In particular, 6G will revolve around a new vision of {\em ubiquitous AI}, a hyper-flexible architecture that brings human-like intelligence into every aspect of networking systems.

\begin{figure*}
\centering
\includegraphics[width=5.6 in]{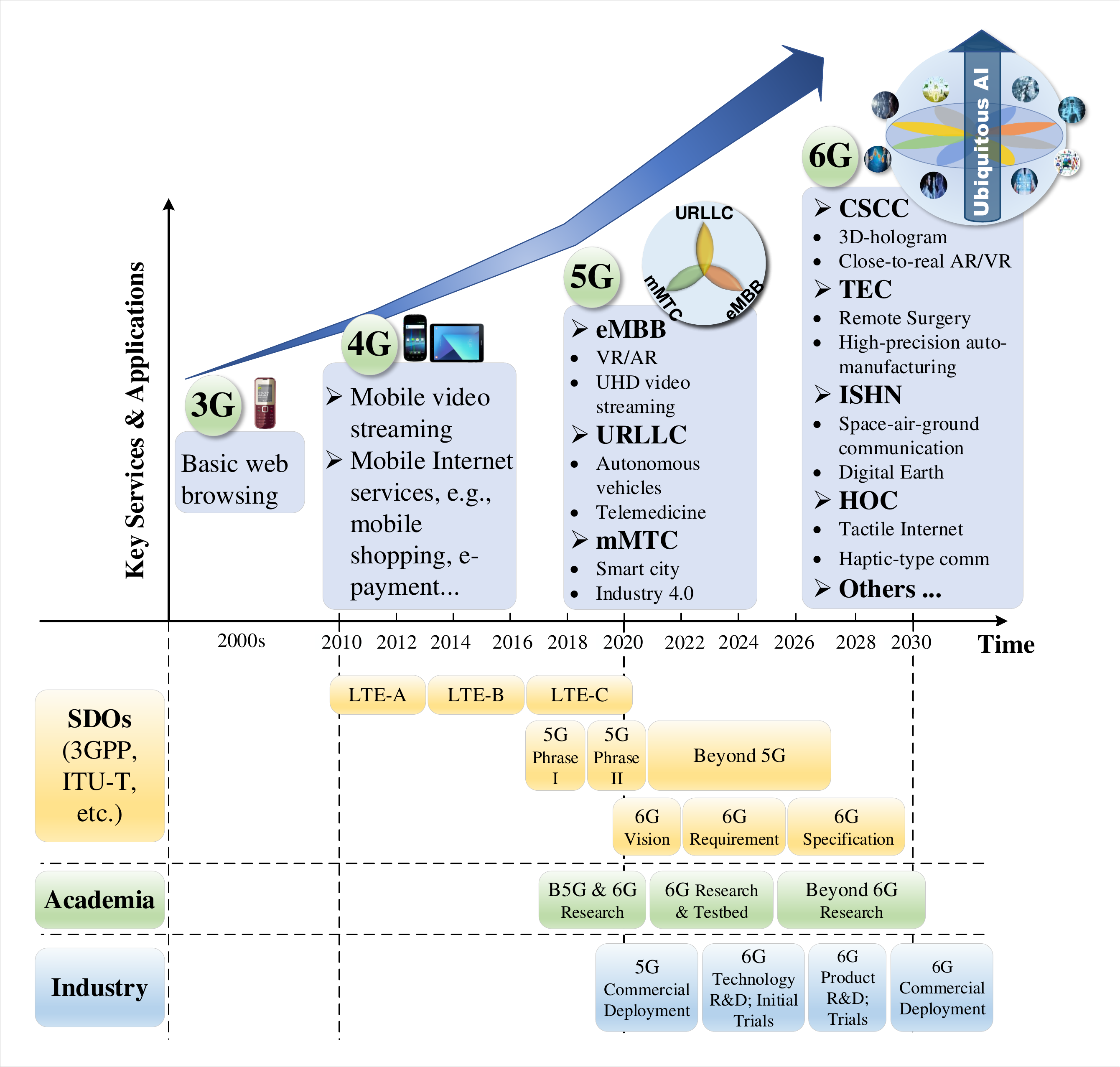}
\caption{Key services, applications, and roadmap for 6G. }
\vspace{-0.2in}
\label{Figure_5G6Gevolution}
\end{figure*}

AI has already proved essential in many applications that influence people's everyday lives. For example, computer-vision-based facial recognition has made mobile shopping and e-payments a reality. Speech recognition and natural language processing-enabled virtual assistants have been applied widely in mobile phones, smart home applicants, and smart vehicles. The potential of AI in optimizing and improving wireless networking systems is already well-recognized\cite{Stanford2016AISurvey,Stoica2017BerkeleyAI}. ITU-T has also established multiple focus groups (FGs), including FGs on machine learning for future networks that include 5G (FG-ML5G) 
and an FG on data processing and management to support IoT and smart cities \& communities (FG-DPM), 
promoting data-driven AI applications in future network system development. There are also initiatives in both academia and industry to develop AI-inspired automated algorithms for improving the efficiency of next-generation networking systems, e.g., 6Genesis project\footnote{https://www.oulu.fi/6gflagship}
led by the University of Oulu in Finland and AT\&T and Microsoft's recent strategic alliance to deliver innovation with cloud computing, AI, and 5G. The key performance requirements and features of 6G suggested by ITU and some authors\cite{ITU2019Network2030, Dang20206GSurvey, Letaief2019Roadmapto6G, Tariq20196GSurvey, Saad20196GSurvey} compared with 5G are summarized in Table \ref{Tabel_5G6GCompare}. The roadmap for 6G evolution according to ITU are described in Figure \ref{Figure_5G6Gevolution}. 

One of the key challenges in designing an AI-based architecture for practical networking systems is how to implement distributed data processing and learning across a massive number of heterogeneous devices. Federated learning (FL) is an emerging distributed AI solution that enables data-driven AI and machine learning (ML) on a large volume of decentralized data that reside on mobile devices\cite{McMahan2017FLfirstpaper}. FL has attracted significant interest due to its ability to perform model training and learning on heterogeneous and potentially massive networks, while keeping all the data localized. Although it is still in the early stages of development, FL-inspired distributed architectures have already been recognized as one of the most promising solutions to fulfill 6G's vision of ubiquitous AI\cite{Li2019FLSurvey,Letaief2019Roadmapto6G}.

\begin{figure}
\centering
\includegraphics[width=3.5 in]{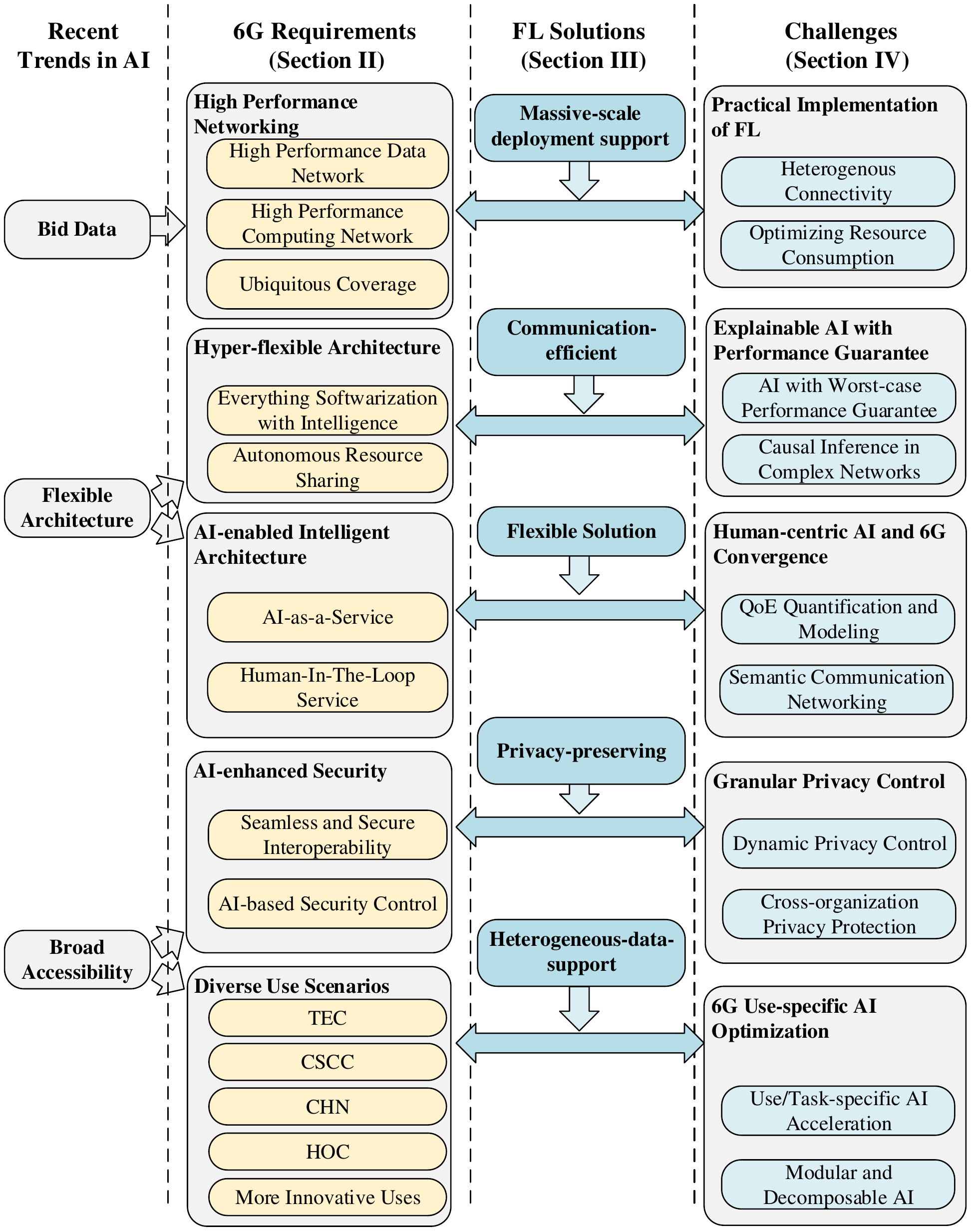}
\caption{Key requirements for 6G and AI as well as possible FL-enabled solutions and future challenges. }
\label{Figure_6GAIFLChallengeRelation}
\vspace{-0.2in}
\end{figure}

In this article, we identify the key requirements and future trends that will drive ubiquitous AI for 6G, especially from the perspective of FL. We propose an FL-based architecture and discuss its potential in addressing key challenges in 6G. To the best of our knowledge, this is the first work that surveys FL and its possible application in 6G systems.  The rest of this article is organized as follows. In Section \ref{Section_Requirements}, we describe the basic requirements for 6G systems, driven by recent trends in AI. 
An FL-enabled ubiquitous AI architecture  as well as its possible applications in 6G are discussed in Section \ref{Section_FL}. Finally, in Section \ref{Section_ChallengesandResearchTopics}, we describe the main challenges and important open research topics in FL-based 6G and conclude the article in Section \ref{Section_Conclusion}. The main structure of this article is summarized in  Figure \ref{Figure_6GAIFLChallengeRelation}.

\begin{table}[tbp]
\centering
\caption{Comparison of 5G and 6G}
\vspace{-0.1in}
\label{Tabel_5G6GCompare}
\scriptsize
\begin{tabular}{|l|l|l|}
\hline
 & \makecell[c]{5G} & \makecell[c]{6G} \\
\hline
\makecell[l]{Network\\ Performance} & \makecell[l]{$\bullet$ Up to 20 Gb/s data rate\\$\bullet$ 1 ms latency \\ $\bullet$ 10-100X energy  \\ \;\; efficiency compared\\
\;\; to 4G} & \makecell[l]{$\bullet$ Up to 1 Tb/s data rate\\$\bullet$ 10-100 $\mu$s latency \\ $\bullet$ 10-100X energy efficiency \\ \;\;\;\;compared to 5G} \\
\hline
Architecture & Service-based architecture &
\makecell[l]{Hyper-flexible architecture\\ based on ubiquitous AI} \\
\hline
Spectrum & 2-72 GHz mmWave band & \makecell[l]{2-300 GHz mmWave\\ and THz band}    \\
\hline
\makecell[l]{Use\\ Scenarios} & eMBB, mMTC, URLLC & \makecell[l]{$\bullet$ Communication Services with\\ \;\; Complex Constraints (CSCC)\\ $\bullet$ Time-engineered\\
\;\; Communications (TEC) \\$\bullet$ Integrated Services involving \\
\;\; Heterogeneous Networks (ISHN) \\ $\bullet$ Human-oriented Communication\\
\;\; Service (HOS) \\ $\bullet$ Others \ldots} \\
\hline
Privacy & \makecell[l]{Improved authentication \& \\ access control over 4G} & \makecell[l]{New AI-based privacy protection \& \\ control framework} \\

\hline
\end{tabular}
\vspace{-0.1in}
\end{table}
\normalsize

\section{6G Requirements and Use Cases} 
\label{Section_Requirements}
Although there is still no consensus on what 6G entails, it is commonly believed that existing progress and evolution of AI will play a dominant role in shaping some of the key requirements of 6G\cite{Letaief2019Roadmapto6G, Dang20206GSurvey}.
The recent success of AI has been made possible mostly by three emerging trends\cite{Stoica2017BerkeleyAI}: (1) {\em big data}, generated by the popularity of mobile Internet and smart services; (2) {\em flexible software and scalable architecture}, inspired by advances in AI algorithms and methodology; and (3) {\em broad accessibility} of affordable services offered by service providers through the mobile Internet. These trends will continue to dominate the future evolution of AI and will form the fundamental enabler of 6G. 
In particular, 6G needs to meet the following five requirements.

\subsection{High Performance Networking}

\subsubsection{High Performance Data Network} 6G will continue to push the boundaries of data collection, transportation, and networking performance, empowering innovative data-driven AI applications across a wide variety of industries. More specifically, 6G is expected to support up to 1 Tbps data rate per user, 10-100 $\mu$s end-to-end latency, up to 99.9999\% end-to-end reliability, and around 10-100 times improvement in energy efficiency over the existing 5G network. In addition, 6G will support highly diversified data types in forms of text, audio, video, AR/VR, haptic-type communication, 3D-hologram, etc., providing new ways to explore knowledge that will benefit various aspects of human activity.

\subsubsection{High Performance Computing Network} 6G will be much more than just a data transportation network. It will also offer ubiquitously available in-network data processing capability to support on-demand computation-intensive and mission-critical applications, such as high-precision remote medical diagnosis and treatment. Massive-scale distributed computational networking has recently been advocated by both industry and academia. In particular, fog computing\footnote{Another concept that is very similar to fog computing is mobile edge computing. In this article, we follow the commonly adopted terminology and define fog computing as a more generalized version of mobile edge computing to include cloud, edge, clients, and things.}---a distributed computing architecture in which data processing and caching are performed by low-cost fog computing servers that are deployed at the edge of the network---has been recently promoted by major telecommunication operators and information communication technology (ICT) companies\cite{XY2018TactileInternet}. Unfortunately, most existing approaches implement fog servers as simple add-ons to the existing networking infrastructure, e.g., a mini-computer attached to a 5G base station (gNB). Also, each fog server can only support a very limited number of preset services, constrained  by  the installed hardware/software platforms and solutions.
A more comprehensive and future-proof architecture that could offer seamless integration of both fog computing and communication network infrastructure  is expected to play a pivotal role in ubiquitous AI-based 6G.

\subsubsection{Ubiquitous Coverage} 6G services and applications will be extended to provide much broader coverage than 5G, integrating terrestrial networks with space-borne/satellite networks that include GEO, MEO, and LEO satellites, air-borne systems incorporating HAPs, LAPs, as well as smaller unmanned aerial vehicles (UAVs)\cite{Kato2019spaceairground}. Underwater communication networks consisting of both commercial and military applications are also expected to be a part of the 6G infrastructure, enabling remote environmental monitoring and control, energy harvesting, and unmanned ocean exploration and mining. Furthermore, the existing terrestrial network will also evolve to a more holistic and agile system that supports ultra-dense indoor deployment and highly dynamic time-critical applications, such as autonomous vehicles and UAVs.

\subsection{Hyper-flexible Architecture}

\subsubsection{Everything Softwarization with Intelligence}
6G is expected to further enhance network softwarization (NS) and network function virtualization (NFV) technologies.
NS and NFV have transformed networking architecture from traditional infrastructure-oriented into software-based, so the physically limited network resource can now be shared by multiple services and users in a highly efficient way. For example, {\em network slicing}, one of the key enabling technologies of 5G, allows common hardware infrastructure to be virtualized into multiple independent logical virtual network functions. Each function can be isolated and tailored as a resource slice to support a specific service. This has sparkled a growing body of research to apply AI techniques, such as reinforcement learning and supervised learning, to optimize in real-time resource distribution among different slices according to demand  dynamics and service requirements\cite{XY2018EHFogComputing}. In addition to the wireless resources, novel technologies will be introduced to support softwarization of a wide variety of other factors that may influence networking performance.  

\subsubsection{Autonomous Resource Sharing}
Similar to earlier wireless generations, 6G will also have to deal with the resource scarcity problem. Although 5G introduced a wide spectrum in 24-72 GHz millimeter wave (mmWave) band for supporting up to 20 Gb/s data rates for bandwidth-hungry applications, these bands tend to experience high path loss and are sensitive to blockage.
6G will continue to explore spectrum at higher frequencies, especially in 80-300 GHz as well as THz bands, with the goal to deliver 1 Tb/s data rates. However, these bands will suffer even more severe propagation loss and inability to penetrate through obstacles. Therefore, developing an AI-based autonomous spectrum sharing mechanism that can dynamically utilize and switch among sub-6 GHz, mmWave, and THz bands according to the service requirements, resource availability, and propagation environment will be a key issue in 6G. In addition to spectrum sharing, novel sharing mechanisms that involve other types of resources should also be explored. 5G has already embraced network sharing as one of the key solutions to address resource scarcity, extending the concept of infrastructure sharing and carrier aggregation within a single operator's network into multi-operator network sharing, which allows two or more operators to share their licensed spectrum as well as unlicensed spectrum with other technologies, such as Wi-Fi. To promote resource sharing between operators, the FCC has recently introduced the concept of ``geographical area licensing" for assigning spectrum licenses in some mmWave bands. According to this concept, operators can auction their spectrum licenses based on geographical locality, which would ultimately drive operators to negotiate and reach 
bilateral/multilateral spectrum leasing/sharing agreements on demand. It is expect that 6G will introduce more creative ways for sharing various resource across networking systems.

\subsection{AI-enabled Intelligent Architecture}
\subsubsection{AI-as-a-Service}
6G is expected to be the first networking technology to offer full-fledged support of AI-as-a-Service (AIaaS), a general framework for delivering AI-oriented services such as data processing, model selection, training, and tuning on demand. In particular, the ability of generating a huge amount of data and performing computation and storage at the edge of the network makes 6G a hotbed for AI applications and service innovation. AIaaS allows users and service providers to directly deploy AI-enabled services without worrying about the details of AI models, parameter tuning, or software/hardware implementation. AIaaS will have the potential to significantly reduce the deployment cost of AI services and improve the resource utilization, catalyzing the widespread adaptation of AI and its enabled services in practical systems.

\subsubsection{Human-In-The-Loop Service}
6G is expected to offer more Human-In-The-Loop (HITL) services, with highly personalized and substantially improved user experience. In particular, new functionality as well as software and hardware solutions are expected to emerge for sensing and detecting novel dimensions of human experience, such as touch, feeling, emotion, and even psychological senses, enabling novel services such as tactile Internet, haptic-type communication, and close-to-real holographic streaming and interaction. These sensed human-relevant data will be learned, aggregated, and jointly synthesized to offer more accurate and highly orchestrated user-specific service experience.

\subsection{AI-enhanced Security}

\subsubsection{Seamless and Secure Interoperability}
Currently, a bewildering array of incompatible communication protocols is being offered for connecting mobile devices to the Internet. Although many of these protocols share a similar technology, e.g., OFDMA, they vary considerably in their implementations, control mechanisms, and operational and accessible resources (e.g., frequency bands). A common interface and mechanisms to allow coordination and interoperation between different wireless protocols are still lacking. 6G is expected to offer smooth convergence between wireless technologies with overlapping use scenarios as well as seamless interoperability among various networking systems covering different dimensions. In particular, 6G is expected to include some very complex networking systems such as smart city and digital Earth, in which many wired and wireless technologies will need to harmoniously coexist, coordinate, and integrate with each other to fulfill common or highly relevant tasks. 6G will also promote more efficient incentive mechanisms and privacy protection solutions to enable fully automatic operations and control across different technologies and systems, including Wi-Fi, satellite (e.g., DAMA), underwater communication (e.g., JANUS) as well as commercial and military systems.

\subsubsection{AI-based Security Control}
With more personalized data and services being offered by 6G, unprecedented amounts of business and personal data will be generated, transported, and processed. These highly sensitive data will make 6G more likely to be exposed to a much diverse forms of security threats than previous generations of wireless technology. A substantially more intelligent solution for security protection and control from individual user device level all the way to networking level must be introduced and enforced.

\subsection{Diverse Use Scenarios}
\label{Section_ServiceFL6G}

Compared to its predecessors, 5G has already taken a major paradigm shift from communication-oriented architecture to service-based architecture (SBA). This shift was driven by the diverse needs of three major use scenarios: enhanced Mobile Broadband (eMBB), Ultra Reliable Low Latency Communications (URLLC), and massive Machine Type Communications (mMTC). 6G will take a more holistic approach, driven by the novel needs of other verticals.
In particular, according to ITU-T and some authors from both academia and industry\cite{ITU2019Network2030, Dang20206GSurvey, Letaief2019Roadmapto6G, Tariq20196GSurvey, Saad20196GSurvey}, 6G is expected  to offer support for the following service types.

\subsubsection{Time-engineered Communications (TEC)} In contrast to 5G's  URLLC  which essentially relies on allocating and preserving redundant resources to improve the latency and reliability of data transmission, 6G will offer more innovative solutions to simultaneously support services with a wide range of timeliness requirements of data delivery. For example, protocol or architecture-level optimization approaches are expected to emerge in 6G to further reduce latency.
In addition, more diverse requirements on the timeliness of data delivery are expected to be supported, including in-time, on-time, coordinated service, as well as services requiring a certain confidence/probability on some latency guarantees.  Examples of TEC may include high precision manufacturing automation, remote surgery, instantaneous response to emergency/safty-related situation, synchronized operations, such as drone swarms.

\subsubsection{Communication Services with Complex Constraints (CSCC)} Although IoT and eMBB have been originally intended for totally different use cases, there are already emerging applications promoting convergence between the two. For example, Industry IoT 5.0 enables real-time surveillance of high-precision manufacturing processes supported by a large number of high-definition cameras and high-performance sensors collecting a large volume of data. Another emerging technology, referred to as {\em digital twins}, will also rely on a massive number of high-performance IoT devices to collect real-time data from various sources to create a high-fidelity digital representation of a physical object or environment. CSCC offers a converged solution and can support services with diverse constraints, including bandwidth, latencies, and security levels.

\subsubsection{Integrated Services involving Heterogeneous Networks (ISHN)} 6G is expected to be a highly complex networking system, involving many heterogenous sub-networks and sub-systems that offer different service types and coverage, such as space-land-communication; underwater-space communication.

\subsubsection{Human-oriented Communication Service (HOS)}
6G is expected to offer creative ways for cyber-physical systems and  human-physical world interactions. For example, the Tactile Internet is expected to become a reality, enabling real-time communications of the human physical interaction, including teleoperation, cooperative automated driving, and interpersonal communication.

\section{FL for Ubiquitous AI in 6G}
\label{Section_FL}

\begin{figure}
\centering
\includegraphics[width=3.5 in]{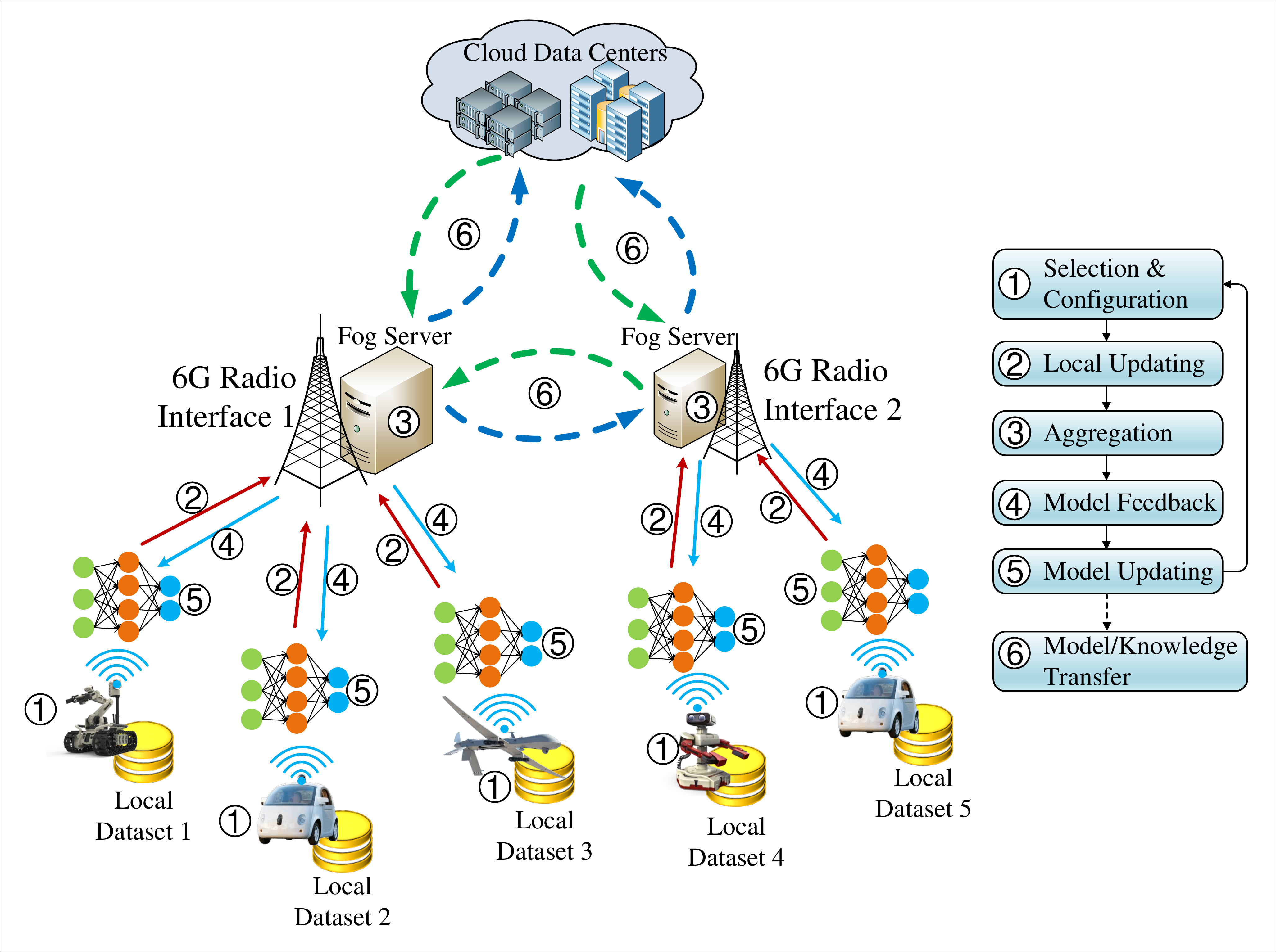}
\vspace{-0.2in}
\caption{FL architecture and detailed procedure. }
\label{Figure_FL}
\vspace{-0.2in}
\end{figure}

\subsection{FL-based Architecture}
In this section, we introduce an FL-based architecture that has the potential to address some of the challenges raised by 6G. In this architecture, a large number of devices associated with different services attempt to jointly construct a common model (image classification, next-word prediction, voice recognition, anomaly detection, etc.) based on locally collected datasets.

The detailed procedure is described in Figure \ref{Figure_FL}. In each communication round, a device will first evaluate its service request, demands, and connectivity conditions, and decide whether to register with the closest fog server via a wired or wireless connection (e.g., a 6G radio interface) to join the training of shared models. Each fog server will then carefully select a subset of devices to participate in this round of training and reject the rest of the registered devices. The fog server will also send back detailed configuration information, such as data structure, sharing state, and model parameters, to each selected device. Each participating device will then perform local computations based on the received configuration information and its local dataset. Updates will be sent from all selected devices to the fog server for aggregation. Once the fog server receives enough updates, it will send the aggregation result back to participating devices. The devices will update their respective models with the received results. The above process will be repeated in the next round with a newly selected subset of devices until the trained model converges or the stopping criteria are met. The learned model as well as the knowledge related to the model can be transferred and utilized by other servers as well as the cloud data center for the benefit of others and/or coordination with other services and users across a wider geographic area\cite{Qiang2019FL}.

\subsection{FL for 6G}
FL has potentials to address some key challenges of 6G.
\subsubsection{FL for Communication-efficient AI} Since the raw data in each device need not to be uploaded to a centralized location, the total traffic transmitted on the network will be reduced. Recent works show that the communication overhead of FL can be further reduced by compressing the data size of each device-side updating or reducing the model update frequency, e.g., by avoiding updates from devices with limited contributions to model convergence\cite{Li2019FLSurvey}.

\subsubsection{FL for Heterogeneous Data Support} The statistical features of datasets collected by different devices can vary significantly due to their different service types, use scenarios, user preferences, etc. FL supports model training among devices with non-identically distributed (non-IID) datasets\cite{Bonawitz2019FLMassiveDeploy}.

\subsubsection{FL for Privacy Preservation}
The updated information uploaded by each device is inspectable but yet cannot be used to recover any useful information about the local dataset. The security level of FL can be further improved by adopting more advanced encryption and security measures, such as  secure aggregation at each fog server and user-level differential privacy\cite{Li2019FLSurveyPrivacy}.

\subsubsection{FL for Massive-scale Deployment Support} FL supports massive-scale network deployment\cite{McMahan2017FLfirstpaper}. In particular, experimental results show that FL can converge to the optimal solution even when the number of devices participating in model training is much larger than the average number of samples in the dataset of each device\cite{Bonawitz2019FLMassiveDeploy}.

\subsubsection{FL for Flexible Networking Architecture} The applicable areas and achievable solutions of FL have been significantly enriched by recent developments on the integration of other state-of-the-art AI solutions, such as deep learning\cite{McMahan2017FLfirstpaper}, reinforcement learning, 
 transfer learning, and generative learning. 

\section{Challenges and Open Research Topics}
\label{Section_ChallengesandResearchTopics}

\subsection{Practical Implementation of FL}

\subsubsection{Heterogenous Connectivity} Although recent experimental results show that it is unnecessary for every device to update the server in every round of model training, FL can only converge to an unbiased solution if all the devices are equally likely to participate in the model training updates. In practical networking systems, however, mobile devices can experience frequent disconnection and decide to leave and join the training process due to the change of interest or service demands. This may lead to an inferior model or biased training results for FL. How to develop a simple mechanism to detect and keep track of the connectivity status of all mobile devices during model training, and how to adjust the bias of the resulting model are still open issues.

\subsubsection{Optimizing Resource Consumption}
The performance of FL is closely related to the availability and reliability of network connectivity as well as the computational capability of both servers and devices. In addition, the communication and computing resource consumption can vary substantially when being applied with different AI algorithms. How to quantify various resource consumption when being applied to different network topologies and services under different scenarios is still an open problem.

\subsection{Explainable AI with Performance Guarantees}

\subsubsection{AI with Worst-case Performance Guarantees}
Most modern networking systems offer worst-case performance guarantees, that is, the network can still offer some supported services even when the worst-case scenarios happen. Typically, these worst-case scenarios are assumed to be known and therefore the impact on various components of the networking system in these scenarios can be pre-evaluated. Unfortunately, 6G is expected to a very complex networking system, where  the possible consequences and scenarios are
much more difficult to measure and calculate. How to develop an explainable AI solution that can be used to calculate and evaluate network performance under different scenarios, and more importantly, offer worst-case performance guarantees is still an open problem.

\subsubsection{Causal Inference in Complex Networks}
In 6G systems, a large number of AI-enabled networking components will be expected to coexist and work together. One of the key challenges for maintaining and debugging such a complex system is to ensure that the network supports {\em causal inference}, i.e., the ability to identify the features or attributes of the input that result in a particular output. The system developer is then able to simulate and faithfully replay the process when evaluating and debugging the network. How to develop a causal inference-enabled AI, especially for the complex networking scenarios of 6G, is still a very challenging task.

\subsection{Human-centric AI and 6G Convergence}

\subsubsection{Quality-of-Experience (QoE) Quantification and Modeling}
6G will be more focusing on optimizing and improving users' QoE than their Quality-of-Service (QoS). QoE
is more related to users' subjective performance when experiencing a specific service. It is not only affected by service-related hardware and software performance, but also influenced by real feeling of the human being, e.g., how much satisfied or unsatisfied the user is.
QoE also depends on users' age, gender, personality, as well as the time, location, and physical environment of the particular service.
It is still unclear how such factors can be combine in model formulation to accurately quantify and evaluate the QoE of human users.

\subsubsection{Semantic Communication Networking}
It is expected that 6G will be transforming from the data or service-oriented networking into semantic networking, a more universal communication framework in which different components of networking system can interact and communicate based on the meaning of the message. In this framework, users and devices with different backgrounds, languages, and protocols will be able to communicate with each other in a highly efficient way.

\subsection{Granular Privacy Control}

\subsubsection{Dynamic Privacy Control}

6G is expected to offer privacy protection and control at a more diverse level and finer granularity, as privacy constraints may differ across different services and/or 
different time and location. For example, an autonomous vehicle needs to mainly focus on the road shape and conditions when driving on a highway, but will need to coordinate its driving behavior with the road-side units and other neighboring vehicles at the intersections or round about. Unfortunately, FL only preserves privacy at the local level by keeping data within each individual device. How to develop novel 6G and AI solutions to offer more controllable (service-specific or even temporal and spatial-specific) and granular  privacy restrictions is an open problem.

\subsubsection{Cross-organization Privacy Protection}
As mentioned earlier, 6G will collect highly sensitive data that can be used to detect and infer various aspects related to  user experience. This data can be generated, transported, and processed by different devices, networking components, and servers associated with different organizations. In other words, 6G must adopt a more holistic data protection mechanism that will be able to protect all the data across different organizations for each specific service.

\subsection{6G Use-specific AI Optimization}

\subsubsection{Use/Task-specific AI Acceleration}
It is well-known that the rate of data generation and consumption have already exceeded  data processing and storage capacity.
The total amount of data will soon exceed the processing and storage capacity of all the hardware infrastructure offered by cloud and fog service providers.
It is, therefore, important to develop some use-specific hardware and software solutions that can accelerate the processing speed and compress the storage space for some common use scenarios or tasks with reduced energy consumption.

\subsubsection{Modular and Decomposable AI}
Modularity is an effective solution that has been adopted by many practical industrial systems to improve the deployment speed and utilization of software and hardware resources. In particular, by decomposing a complex system into a set of flexible and reusable components, the same set of modules and solutions can be dynamically deployed and adjusted for different algorithms and solutions. Designing modular and decomposable AI solutions with generic AI modules and algorithmic solutions that can be reused for some common or similar 6G use scenarios and tasks will significantly accelerate the adaptation of AI in 6G.

\section{Conclusion}
\label{Section_Conclusion}

This article has provided an overview of a possible research roadmap for 6G, from the perspective of AI. Potential requirements and challenges of 6G and AI convergence have been identified. An FL-based architecture that supports distributed data analysis and training at massive scales of mobile devices has been proposed. We discussed the possible challenges that can be addressed by the proposed architecture and presented open research problems for FL-enabled 6G. We hope this article will spark further interest and open new research directions into the evolution of FL and its applications towards ubiquitous AI in 6G.

\section*{Acknowledgment}
We would like to thank Prof. Ross Murch at the Hong Kong University of Science and Technology and Prof. Luiz DaSilva at the CONNECT, Trinity College Dublin, for useful comments on this manuscript.

\bibliography{DeepLearningRef}
\bibliographystyle{IEEEtran}

\begin{IEEEbiographynophoto}{Yong Xiao}(S'09-M'13-SM'15) is a professor in the School of Electronic Information and Communications at the Huazhong University of Science and Technology (HUST), Wuhan, China. His research interests include machine learning, game theory, and their applications in cloud/fog/mobile edge computing, green communication systems, wireless networks, and Internet-of-Things (IoT).
\end{IEEEbiographynophoto}

\begin{IEEEbiographynophoto}{Guangming Shi} is a professor in the School of Artificial Intelligence, the Xidian University. He has authored or co-authored more than 200 research papers. His research interests include compressed sensing, theory and design of multirate filter banks, image denoising, low bit-rate image/video coding, the implementation of algorithms for intelligent signal processing (using DSP and FPGA), deep neural networks, and artificial intelligence.
\end{IEEEbiographynophoto}

\begin{IEEEbiographynophoto}{Marwan Krunz}(S'93-M'95-SM'04-F'10) is the Kenneth VonBehren Endowed Professor in the Department of Electrical and Computer Engineering, University of Arizona. He is the center director for BWAC, a multi-university NSF/industry center that focuses on next-generation wireless technologies and applications. His research emphasis is on resource management, network protocols, and security for wireless systems.
\end{IEEEbiographynophoto}

\end{document}